\documentstyle[editedvolume]{crckapb} 
\input epsf

\def\IUE{{\it IUE}}
\def\HST{{\it HST}}
\def\deg{\ifmmode ^{\rm o} \else $^{\rm o}$\fi}
\def\cc{\ifmmode {\rm cm}^{-3} \else cm$^{-3}$\fi}
\def\vFWHM{\ifmmode v_{\mbox{\tiny FWHM}} \else
            $v_{\mbox{\tiny FWHM}}$\fi}
\def\arcsec{\ifmmode '' \else $''$\fi}
\def\arcmin{\ifmmode ' \else $'$\fi}
\def\arcsecpoint{\ifmmode ''\!. \else $''\!.$\fi}
\def\arcminpoint{\ifmmode '\!. \else $'\!.$\fi}
\def\micron{\ifmmode \mu{\rm m} \else $\mu$m\fi}
\def\kms{\ifmmode {\rm km\,s}^{-1} \else km\,s$^{-1}$\fi}
\def\Hubble{\ifmmode {\rm km\,s}^{-1}\,{\rm Mpc}^{-1} 
        \else km\,s$^{-1}$\,Mpc$^{-1}$\fi}
\def\ergsec{\ifmmode {\rm ergs\;s}^{-1} \else ergs s$^{-1}$\fi}
\def\ergscm{\ifmmode {\rm ergs\,s}^{-1}\,{\rm cm}^{-2}
          \else ergs\,s$^{-1}$\,cm$^{-2}$\fi}
\def\ergscmA{\ifmmode {\rm ergs\,s}^{-1}\,{\rm cm}^{-2}\,{\rm \AA}^{-1}
          \else ergs\,s$^{-1}$\,cm$^{-2}$\,\AA$^{-1}$\fi}
\def\ergscmHz{\ifmmode {\rm ergs\,s}^{-1}\,{\rm cm}^{-2}\,{\rm Hz}^{-1}
          \else ergs\,s$^{-1}$\,cm$^{-2}$\,Hz$^{-1}$\fi}
\def\Msun{\ifmmode M_{\odot} \else $M_{\odot}$\fi}
\def\Lsun{\ifmmode L_{\odot} \else $L_{\odot}$\fi}
\def\qo{\ifmmode q_{0} \else $q_{0}$\fi}
\def\Ho{\ifmmode H_{0} \else $H_{0}$\fi}
\def\ho{\ifmmode h_{0} \else $h_{0}$\fi}
\def\qo{\ifmmode q_{0} \else $q_{0}$\fi}
\def\ao{\ifmmode a_{0} \else $a_{0}$\fi}
\def\to{\ifmmode t_{0} \else $t_{0}$\fi}
\def\ltsim{\raisebox{-.5ex}{$\;\stackrel{<}{\sim}\;$}}
\def\gtsim{\raisebox{-.5ex}{$\;\stackrel{>}{\sim}\;$}}
\def\Halpha{\ifmmode {\rm H}\alpha \else H$\alpha$\fi}
\def\Hbeta{\ifmmode {\rm H}\beta \else H$\beta$\fi}
\def\Hgamma{\ifmmode {\rm H}\gamma \else H$\gamma$\fi}
\def\Hdelta{\ifmmode {\rm H}\delta \else H$\delta$\fi}
\def\Lya{\ifmmode {\rm Ly}\alpha \else Ly$\alpha$\fi}
\def\Lyb{\ifmmode {\rm Ly}\beta \else Ly$\beta$\fi}
\def\hi{\ifmmode \mbox{{\rm H}\,{\sc i}} \else H\,{\sc i}\fi}

\def\heii{He\,{\sc ii}}

\def\ciii{\ifmmode {\rm C}\,{\sc iii} \else C\,{\sc iii}\fi}
\def\civ{\ifmmode {\rm C}\,{\sc iv} \else C\,{\sc iv}\fi}

\def\nv{N\,{\sc v}}

\def\o5007{[O\,{\sc iii}]\,$\lambda5007$}
\def  \Rin         {\hbox{$ {R_{\rm in}} $}}
\def  \Rout        {\hbox{$ {R_{\rm out}} $}}
\def  \RBLR        {\hbox{$ {R_{\rm BLR}} $}}
\def  \Vin         {\hbox{$ {V_{\rm in}} $}}

\def  \nh          {\hbox{$ {n_{\rm H}} $}}      
\def  \Ncol        {\hbox{$ {N_{\rm col}} $}}      
\def  \kms         {\hbox{km s$^{-1}$}}          
\def  \cc          {\hbox{cm$^{-3}$}}

\def  \mic         {$\mu$m}

\def  \etal        {{\rm et al.}}
\def  \La          {\ifmmode {\rm Ly}\alpha \else Ly$\alpha$\fi}
\def  \Ka          {\ifmmode {\rm K}\alpha \else K$\alpha$\fi}
\def  \Lb          {\ifmmode {\rm L}\beta \else L$\beta$\fi}
\def  \Ha          {\ifmmode {\rm H}\alpha \else H$\alpha$\fi}
\def  \Hb          {\ifmmode {\rm H}\beta \else H$\beta$\fi}
\def  \Pa          {\ifmmode {\rm P}\alpha \else P$\alpha$\fi}
\def  \CIIIb       {\ifmmode {\rm C}\,{\sc iii]}\,\lambda1909
                     \else C\,{\sc iii]}\,$\lambda1909$\fi}
\def  \CIV         {\ifmmode {\rm C}\,{\sc iv}\,\lambda1549
                     \else C\,{\sc iv}\,$\lambda1549$\fi}
\def  \MgII         {\ifmmode {\rm Mg}\,{\sc ii}\,\lambda2798
                     \else Mg\,{\sc ii}\,$\lambda2798$\fi}
\def  \OVI         {\ifmmode {\rm O}\,{\sc vi}\,\lambda1035
x
                     \else O\,{\sc vi}\,$\lambda1035$\fi}

\begin{opening}
\title{REVERBERATION MAPPING AND THE PHYSICS  OF \protect\\
ACTIVE GALACTIC NUCLEI}
\author{HAGAI NETZER}
\institute{School of Physics and Astronomy and The Wise Observatory,
Tel Aviv University, Tel Aviv 69978, ISRAEL}
\author{BRADLEY M.\ PETERSON}
\institute{Department of Astronomy, The Ohio State University,
174 West 18th Avenue, Columbus, OH  43210, USA}
\end{opening}

\runningtitle{REVERBERATION MAPPING OF AGN}

\begin{document}

\begin{abstract}
Reverberation-mapping campaigns have revolutionized our understanding
of AGN. They have allowed the direct determination of the broad-line
region size, enabled mapping of the gas distribution around the central
black hole, and are starting to resolve the continuum source structure.
This review describes the recent and successful
campaigns of  the International AGN Watch consortium, outlines the
theoretical background of reverberation mapping and the calculation
of transfer functions, and addresses the fundamental difficulties of
such experiments. It shows that such large-scale experiments have
resulted in a 
``new BLR'' which is considerably different from the one we 
knew just ten years ago. 
We discuss in some detail the more important new results, including the
luminosity--size--mass relationship for AGN, 
and suggest ways to proceed in the near future.
\end{abstract}

\section{Introduction}
Continuum variability was one of the earliest recognized characteristics of
quasars, the highest-luminosity active galactic nuclei (AGN). In contrast,
continuum variations were not confirmed in their lower-luminosity
cousins, Seyfert galaxies, until a quarter century after they had been
identified as a separate class of object. Ultimately the shared
characteristic of continuum variability helped establish the link between
quasars and Seyfert galaxies, and in both classes of object
it led immediately to recognition
that large-amplitude, short time-scale variations place strong constraints
on the size of the continuum-emitting region. This is a cornerstone
of the black-hole model of active nuclei. While it has been realized
subsequently that the most violent variability seen in AGN  arises in 
a relativistically beamed component (which dominates the spectra of
the subset of AGN known as ``blazars''), 
the conclusion that the continuum-emitting region is small
is still valid, and indeed the physical scales inferred from
variability are consistent
with thermal emission from accretion disks. The origin of the variations
remains unknown.

Some fifteen years ago was it becoming apparent that
the broad emission lines seen in AGN spectra also vary, both in flux and
in profile. It was established fairly quickly that the emission-line
fluxes vary in response to continuum changes, but the exact time scales
for response remained controversial on account of the relatively few, poorly
spaced observations that had been obtained in even the best-studied cases.
Line-profile variations had been clearly detected in a few cases, and it was
suggested by several authors that these might be attributable to
``excitation inhomogeneities'' resulting from non-uniform illumination
of the broad-line region (BLR) due to light travel-time effects. 
It was widely appreciated that this afforded an opportunity to map out
the spatial and kinematic distribution of the line-emitting clouds:
by carefully following continuum variations and their subsequent
effect on the emission-line fluxes and profiles, one can highly constrain
the phase-space distribution of the line-emitting gas. 
Blandford \& McKee (1982) were the first to articulate the 
mathematical formalism for this process, 
which they called ``reverberation mapping''.  

The potential importance of the reverberation-mapping technique to AGN 
astrophysics is profound: if we can determine the structure and kinematics
of the BLR, it is possible to determine the  effects of various
forces in the immediate vicinity of the central engine 
(gravity and  radiation pressure)
 and under some conditions, to 
determine the mass of the central engine itself, thus possibly testing
the black-hole paradigm directly. The BLR itself is thought to be
comprised of a large number (at least $\sim 10^5$, and by some arguments
more than $10^8$) of individual clouds that have, by nebular standards,
quite high densities   ($n_{\rm e} \gtsim 10^9$\,\cc). These clouds move
highly supersonically, at thousands of kilometers per second, and
cover only about 10\% of the sky as seen from central source.
The BLR is much too small to be resolvable spatially, with an angular
extent of only $\sim10\,\mu$arcsec even in nearby ($cz/H_0 \approx 50$\,Mpc)
Seyfert galaxies.

\section{Reverberation-Mapping Techniques}
\subsection{Fundamentals}
The basic idea of reverberation mapping is similar to that which
underlies, for example, Doppler-weather mapping: the time-delayed,
Doppler-shifted response of a system to a known input signal is
used to infer the structure and kinematics of the responding system.
In the case of reverberation mapping, we passively observe the
input signal, which is generated by the AGN continuum source. In
the absence of reasons to assume otherwise, we make the following
simplifying assumptions:
\begin{enumerate}
\item
 The continuum is supposed to originate in a single, central
source whose UV/optical radiation producing size  
is thought to have a spatial extent of 10--100 gravitational radii, 
$R_{\rm g} \approx 1.5 \times 10^{13}\,(M/10^8\,M_{\odot})$\,cm. 
This estimate is based on accretion-disk models that are generally 
consistent with  the observable parameters.
We note in particular that isotropic emission from the central source
does not have to be assumed.
\item 
The light-travel times across the BLR
 are found to be
in the range of days to weeks. 
The time scale for response of individual BLR clouds to changes in the
ionizing flux is given by the recombination time $\tau_{\rm rec} \approx
(n_{\rm e} \alpha_{\rm B})^{-1}$, where $\alpha_{\rm B}$ is the 
case B hydrogen recombination coefficient. For BLR densities,
$\tau_{\rm rec} \approx 1$\,hr, so the cloud response time is virtually
instantaneous and can be neglected. We must also assume that the
BLR structure and kinematics are constant  over the duration of
the monitoring experiment. This places practical time
limits on the duration of single experiments, as we discuss further
below.
\item 
There is a simple, although not necessarily linear, relationship
between the observed UV or optical continuum and the ionizing continuum
that is driving the line variations. In this context, ``simple'' means
that the continuum variations in these different continuum bands 
appear to be generally similar, without pronounced
differences in the relative structure of the light curves.
\end{enumerate}

The duration of a monitoring campaign $\tau$(campaign)
is never optimal because of the very nature
of the gas distribution and velocity field of the BLR.
As emphasized below, one of the most important new results of the AGN Watch
campaigns is the evidence for a 
``thick'' BLR  geometry in several objects.
This means that the outer edge of the cloud
system  is tens, and perhaps hundreds,
of light days away from the continuum source, while the inner boundary is
only a few light days away. Assuming a  BLR which
extends from \Rin\ to \Rout, the observed line response reflects  
continuum variations that occurred
at times up to $\Rout/c$ before the beginning of the observation. The first
$\Rout/c$ days of the campaign  thus can be of limited use.
On the other hand, the dynamical time of the system at \Rin\
determines the maximum
useful monitoring period since the gas distribution is likely to 
change on this time scale. This time is given by
\begin{equation}
\tau_{\rm dyn} = x R_{\rm out}  / V_{\rm in} 
\end{equation}
where \Vin\ is the typical velocity at \Rin, and  $x=\Rin/\Rout$.
Using $\beta_{\rm in}=\Vin/c$, we find
\begin{equation}
\tau{\rm (campaign)} \approx \frac{R_{\rm out}}{c} 
\left( \frac{x}{\beta_{\rm in}} -1 \right).
\end{equation}
Assuming Keplerian orbits, the velocity in the innermost part is obtained from
the full width at zero intensity (FWZI) 
of the emission lines. Typically,
this is  $\beta_{\rm in} \approx 0.03$. Since $x$ is of 
order $0.1$ or smaller, the useful campaign time
is of the order of $2\Rout/c$ or less, 
i.e., a few months for typical Seyfert 1 luminosities.
Obviously, the gas distribution (and the transfer function, see below)
may be stable
over many dynamical times, but this has not yet  been
established. Thus, multiple-year campaigns are not necessarily 
more useful than few-month campaigns in determining the
BLR gas distribution and velocity field.

\subsection{The transfer function}

Under the assumptions outlined above, the emission-line response as a function
of time and line-of-sight velocity (i.e., Doppler shift) can be
written as 
\begin{equation}
\label{eq:TF}
L(v,t) = \int^{\infty}_{-\infty} \Psi(v, \tau)C(t - \tau)d\tau, 
\end{equation}
where $C(t)$ is the  continuum light curve, and  
$\Psi(v,\tau)$ is the ``transfer function''
which depends on the BLR geometry,  kinematics, and reprocessing physics.
The goal of reverberation-mapping experiments is to provide
light curves $C(t)$ and $L(v,t)$ that can be used to solve for the
transfer function, and infer the characteristic 
of the emission-line cloud system.
A stable and unique solution to an integral equation of this
form requires a large amount of high-quality data, and this
is the practical difficulty in reverberation mapping.
In practice, high-quality spectra of faint objects are not easy to
come by, and as a result most experiments to date have focused on
the somewhat simpler problem of solving for 
the {\em one-dimensional transfer function} $\Psi(\tau)$, which
gives the response of the entire emission line integrated over
line-of-sight velocity, i.e.,
\begin{equation}
\label{eq:1dTF}
L(t) = \int^{\infty}_{-\infty} \Psi(\tau)C(t - \tau)d\tau. 
\end{equation}
Even more often, the actual data are so sparse that all we can 
determine is the first moment of $\Psi(\tau)$;
it is easily shown (Penston 1991; Peterson 1993) that 
convolving eq.\ (\ref{eq:1dTF}) with $C(t)$ yields
\begin{equation}
\label{eq:CCF}
{\rm CCF}(\tau) = \int\Psi(\tau'){\rm ACF}(\tau - \tau') d\tau',
\end{equation}
where CCF$(t)$ and ACF$(t)$ are the line--continuum cross-correlation
function and continuum autocorrelation function, respectively.
Cross-cor\-re\-la\-tion of $C(t)$ and $L(t)$ thus yields 
a first moment of $\Psi(\tau)$, a time scale for emission-line response
that is often referred to as the emission-line ``lag''.

Equations (\ref{eq:TF}) and (\ref{eq:1dTF}) are {\em linear}\/
equations, which seems to introduce another hidden assumption
that is contrary to our previous statement that linearity
does not need to be assumed for the line response.
However, in practice, the transfer equation is solved by replacing
$L(t)$ and $C(t)$ with their difference from the mean values, e.g.,
$\Delta L(t) = L(t) - \langle L \rangle$, which removes the effects of
non-variable components and is equivalent to a first-order expansion
of the transfer equation. Thus, mild nonlinearity does not pose a
problem.

\subsection{An example}

The transfer function introduced earlier gives the 
observed response of an emission line as a function of time delay
and line-of-sight velocity to a delta-function continuum outburst,
as is obvious from eq.\ (\ref{eq:TF}). For illustrative purposes,
we will consider the response of 
a simple and specific, but easily generalizable and 
even possibly relevant, 
BLR model, namely clouds on randomly inclined, circular Keplerian
orbits. Consider first clouds in orbits at inclination $i=90\deg$, i.e.,
with the line of sight in the orbital plane. Positions along the
orbital path are specified by the polar coordinates $r$ and $\theta$ as
defined in Fig.\ 1a. Each position on the orbit projects to a
unique position in velocity--time-delay space, as shown
in Fig.\ 1b. An isotropically emitted  
continuum outburst will be followed by an
emission-line response that is time delayed by the additional path
length to the observer that must be traversed
by (a) the ionizing photons that travel outward from the
central source and are intercepted by BLR clouds and (b) the resulting
emission-line photons that are emitted in the direction of the observer;
such a time-delayed path is shown as a dotted line in Fig.\ 1a.
At some time delay $\tau$, the emission-line response recorded by the
observer will be due to all clouds that lie on a surface of
constant time delay (an ``isodelay surface'') given by
the length of the dotted path in Fig.\ 1a,
\begin{equation}
\tau=(1+\cos\theta)r/c,
\end{equation}
which is the equation of a parabola in polar coordinates. The
intersection of the isodelay surface and the cloud orbit identifies
the clouds that are responding at time delay $\tau$. If both
clouds shown in Fig.\ 1a are moving counterclockwise at orbital 
speed $V_{\rm orb}=(GM/r)^{1/2}$, where $M$ is the mass of the central
source, their observed Doppler-shifted
velocities are $v = \pm V_{\rm orb}\sin\theta$, and thus the locations
of the two clouds project to the two different points in
velocity--time-delay space, as shown in Fig.\ 1b. 
The entire circular orbit is seen to project to an
ellipse in the velocity--time-delay diagram; the zero time-delay
point represents the BLR clouds that lie along the line of sight
(at $\theta = 180\deg$ and line-of-sight velocity $v=0$),
and the largest line-of-sight velocities are measured at $\theta=\pm 90\deg$,
where $\tau = r/c$. The range of time delays extends up to $2r/c$,
corresponding to the response from
the far side of the BLR (i.e., $\theta = 0\deg$).
If we consider identical orbits at lower inclinations, it is easy to
see that the range of time delays decreases from $[0,2r/c]$ to
$[(1 - \sin i)r/c$, $(1+\sin i)r/c]$, 
and line-of-sight velocities similarly decrease by
a factor of $\sin i$. The projection of such an orbit into
velocity--time-delay space is thus an ellipse that has the same
center (at $v=0$, $\tau=r/c$) and ellipticity, but axes that are
smaller by a factor of $\sin i$. For $i = 0$\deg,
the ellipse contracts to a single point at time delay $r/c$ and
line-of-sight velocity $v = 0$. For a system of
clouds in circular orbits of radius $r$ and random inclinations,
the ellipse shown in Fig.\ 1b becomes completely filled in, as shown
in Fig.\ 1c, and this is the transfer function for a system of
clouds in circular Keplerian orbits of radius $r$ and random inclinations.
\begin{figure}
\epsfxsize\hsize
\epsfbox{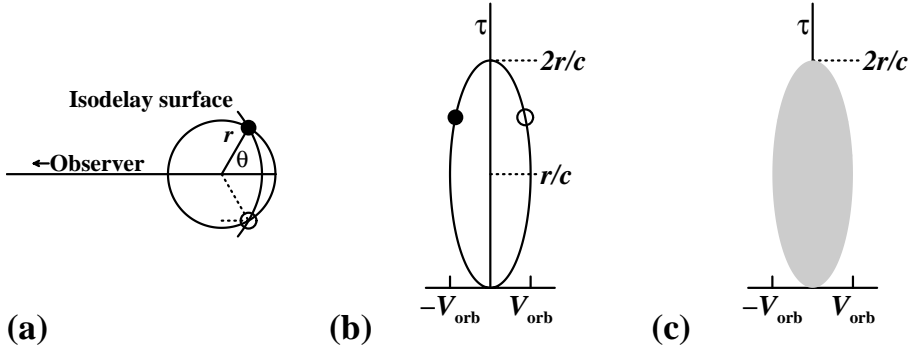}
\caption{(a) In this illustrative model, 
the BLR clouds are distributed along a circular orbit centered
on the central continuum source at inclination $i=90$\deg, with the clouds
orbiting counterclockwise. Emission-line clouds respond to a continuum
outburst with time delay $\tau = (1 + \cos\theta)r/c$, which 
compared with the photons from the central source that travel directly to
the observer, is the additional
path length this signal must travel to the distant observer to the left,
as shown by the dotted line. At the time delay shown, 
two clouds are responding, 
the upper one approaching the observer  and the lower one
receding. (b) The points on the circular orbit in (a) 
project to an ellipse in the velocity--time-delay plane. The
locations of the two clouds in (a) are shown. (c) For circular orbits
at inclinations less than 90\deg, the axes of 
both of the ellipses are decreased
by a factor $\sin i$ and the center remains at $v = 0, \tau =r/c$. 
Thus, for a random distribution
of inclinations, the response of the BLR occurs over the full range
of radial velocities and time delays limited by the
$i=90$\deg\ case.}
\end{figure}

How does the transfer function change if the continuum emission is
not isotropic? To illustrate, we now consider the case where the
continuum radiation is confined to biconical beams of semi-opening angle
$\omega$ at inclination 
$i_{\rm beam}$ to the line of sight, as shown in Fig.\ 2a.
On spatially resolved scales, AGN show such biconical structure, although
it is not at all obvious that this geometry applies on scales
as small as the BLR. 
A biconical beam illuminates the BLR clouds during only {\em parts} of their
orbits, and as shown in Fig.\ 2b, emission-line response is observed only
in certain loci in the velocity--time-delay plane. Again, extending this
to a system of BLR clouds in circular orbits of random inclination,
the ellipses in the velocity--time-delay plane become partially filled in,
as shown in Fig.\ 2c. The effects of varying both $\omega$ and 
$i_{\rm beam}$ are
shown by Goad \& Wanders (1996).
\begin{figure}
\epsfxsize\hsize
\epsfbox{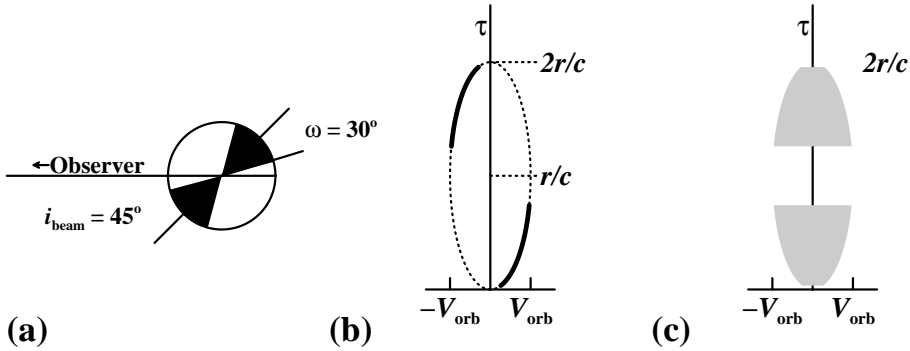}
\caption{(a) The BLR clouds are distributed along a circle as in Fig.\ 1,
but are now illuminated by an anisotropic continuum (shaded) of opening
half-angle $\omega =30$\deg\ and inclined  to the line of sight
by $i_{\rm beam} = 45$\deg.
(b) Since only part of the orbit in (a) is illuminated, only certain
loci in the velocity--time-delay plane show emission-line response.
(c) For a random distribution of {\em orbital}\/ inclinations,
the response of the BLR is similarly localized in time delay. 
There is no response in this case ({\it i}\,) near time delay $r/c$
because the continuum does not illuminate any material in orbits
with $i \approx 0$\deg\ or ({\it ii}\,) near $\tau = 0$ or $\tau =2r/c$ since
the material along the light of sight is out of the beam.}
\end{figure}

It is straightforward to consider more complex models, especially if
the response can be assumed to be approximately linear, in which case
transfer functions for complex BLR geometries and continuum anisotropies
can be constructed by addition of properly weighted simple transfer functions.
For example, partial continuum anisotropy can be modeled as a sum of 
isotropic and anisotropic geometries. Similarly, the response of
a disk can be modeled by a
summation of transfer functions for circular Keplerian orbits of
various $r$ and fixed $i$. The transfer function for a thick spherical
shell (as in Figs.\ 1c and 2c) can be constructed by adding together
transfer functions for thin spherical shells of varying radius; note that
as $r$ increases, the velocity--time-delay 
ellipses become taller and narrower, as the major axis is
proportional to $r$ and the minor axis decreases like $r^{-1/2}$.

In Fig.\ 3, we 
show as an example a thick-shell model that is an extension
of the thin-shell model in Fig.\ 2, and uses the same continuum beaming
parameters. We chose this particular model for two reasons: first,
both the one-dimensional transfer function and the
variable part of the line profile are double peaked. This makes the
important point that such structures are {\em not}\/ unambiguous
signatures of rotating disks or biconical flows (e.g., Welsh \& Horne 1991;
P\'{e}rez, Robinson, \& de la Fuente 1992). Second, this particular
model seems to be grossly {\em consistent}\/
with the observed transfer function
in one of the best-studied cases, the \civ\ line in NGC 5548,
as we will discuss below.
\begin{figure}
\epsfxsize\hsize
\epsfbox{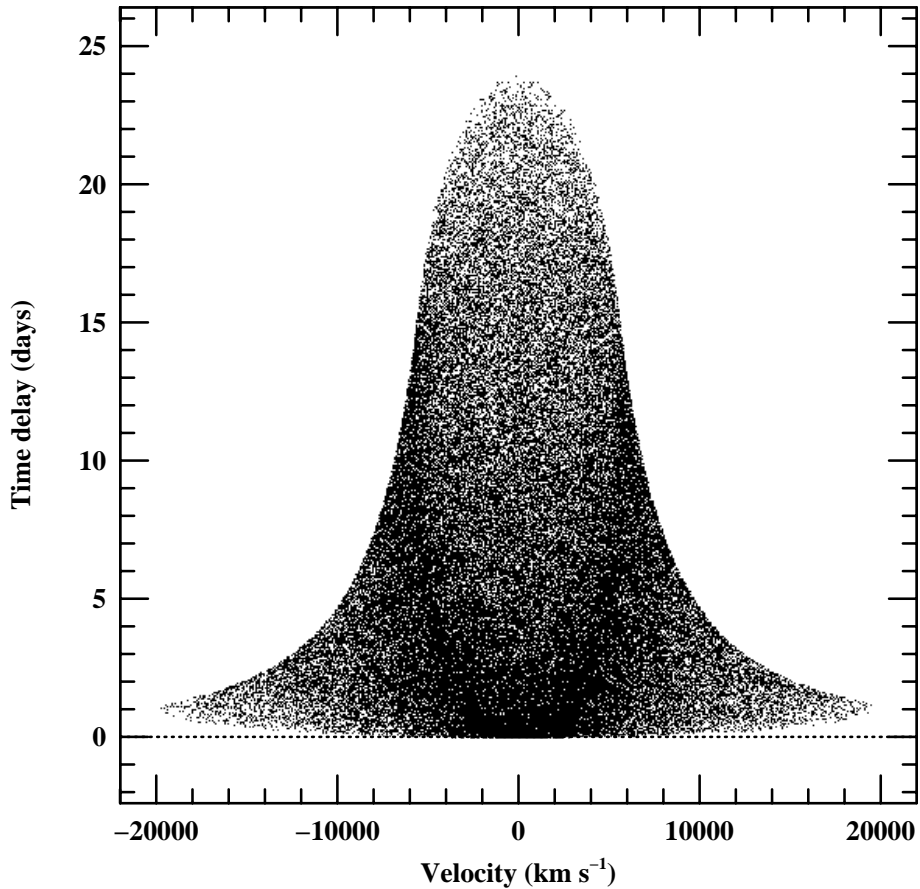}
\caption{A model two-dimensional (i.e., response as a function
of both line-of-sight velocity and time delay) transfer function.
This model is consistent with 
the \CIV\ response in NGC~5548 (Wanders et al.\ 1995),
based on \HST\/ and \IUE\/ data from Korista et al.\ (1995).}
\end{figure}

We close this part of the discussion by reminding the reader 
that the transfer function does {\em not} give a model-independent
six-dimensional map of the BLR in phase space; the velocity-dependent
response of an emission line localizes the gas only to an
isodelay surface. The transfer function does, however,
provide a strong constraint on the phase-space distribution
of the BLR gas, and one must
test BLR models by computing transfer functions for various lines
and comparing these with the observations. It is also important to
remember that the transfer function maps out  the {\em responsivity}\/
of the emission-line gas. We shall come back to this point in \S{4}.

\section{Reverberation-Mapping Experiments}
Equations (\ref{eq:TF}) and (\ref{eq:1dTF}) are examples
of one of the most common types
of problems encountered in physics, a convolution integral with 
an unknown Green's function, which we here call the transfer function.
The observational goal of reverberation-mapping experiments is to
use the light curves $C(t)$ and $L(v,t)$ to solve for the transfer
function and then use this to test directly various models of the BLR.
What makes the reverberation problem different from other applications
is that the sampling of $C(t)$ and $L(v,t)$ is nearly always irregular,
limited in both temporal resolution and duration,
and the data are often quite noisy and sometimes plagued by systematic
errors. These limitations have led to development of specialized
methodologies for time-series analysis. 
The obvious method of Fourier inversion (which was
what was originally suggested by Blandford \& McKee 1982) performs 
poorly on account of the limitations listed above. Better methods include
the maximum entropy method (MEM; Horne 1994, and this volume),
the SOLA method (Pij\-pers \& Wanders 1994), and
regularized linear inversion (Krolik \& Done 1995). 
Even cross-correlation techniques (eq.\ \ref{eq:CCF}) have been
specially adapted to these programs (see White \& Peterson 1994 for
a critical comparison of widely used methods, and Alexander's paper in this
volume for discussion of a new method).

As mentioned earlier, UV and optical spectroscopic monitoring of
a few Seyfert galaxies during the 1980s constituted
``proof-of-concept'' for reverberation mapping; it was clear that
the lines did indeed respond to continuum variations, on surprisingly
short time scales (see Peterson 1988 for a summary of these early
programs). The next step was to acquire suitable time series to measure
the response time scales accurately, i.e., determine the lags for
various emission lines (eq.\ \ref{eq:CCF}). As a practical matter,
monitoring programs
to achieve limited goals can be carried out at individual observatories.
This has been demonstrated convincingly by ground-based results obtained
at the Wise Observatory (e.g., Maoz et al.\ 1990, 1991; Netzer et al.\ 1990)
at Ohio State University (e.g., Peterson et al.\ 1993; 
Kassebaum et al.\ 1997; Peterson et al.\ 1997),
at CTIO (Winge et al.\ 1995, 1996), and 
at La Palma (by the ``LAG'' collaboration, whose work has been 
well summarized by Robinson 1994).
However, probably the greatest success has been been achieved by the large 
consortium known as the ``International AGN Watch'' (Alloin et al.\ 1994),
which has carried out several multi-wavelength monitoring programs that
have been anchored by UV spectroscopy with 
the {\it International Ultraviolet Explorer (IUE)}\/ and
the {\it Hubble Space Telescope (HST)}\/ 
and optical spectroscopy with a large
network of ground-based telescopes. The International AGN Watch
efforts have included two major UV campaigns on NGC 5548
(Clavel et al.\ 1991; Korista et al.\ 1995) complemented by a continuing
ground-based effort (Peterson et al.\ 1991, 1992, 1994;
Dietrich et al.\ 1993; Romanishin et al.\ 1995), and
similar campaigns on NGC 3783 (Reichert et al.\ 1994; Stirpe et al.\ 1994),
Fairall 9 (Rodr\'{\i}guez-Pascual et al.\ 1997; Santos-Lle\'{o} et al.\ 1997),
and 3C~390.3 (O'Brien et al.\ 1997; Dietrich et al.\ 1997). Other
monitoring programs have been built around these projects, including
extreme UV monitoring of NGC 5548 with 
the {\it Extreme Ultraviolet Explorer (EUVE)}\/
(Marshall et al.\ 1997),
a multi-wavelength snapshot of NGC 3783 (Alloin et al.\ 1995),
and long-term X-ray monitoring of 3C 390.3 with {\it ROSAT}
(Leighly et al.\ 1997). A purely
ground-based campaign was also carried out on Mrk 509 (Carone et al.\ 1996).
This group also undertook an intensive multi-wavelength program on NGC 4151
(Crenshaw et al.\ 1996; Kaspi et al.\ 1996a; Warwick et al.\ 1996;
Edelson et al.\ 1996), although the limited duration of this experiment
(about 10 days) precluded learning much about the emission-line response
in this object.

The results on NGC 5548 (Korista et al.\ 1995)
and NGC 4151 (Ulrich \& Horne 1996) represent the state of the art
in reverberation mapping. The transfer functions are not well-determined,
as they are based on $\ltsim50$ data points that are not noise-free.
However, these two results are not likely to remain the last word on
reverberation mapping for long: at the time of writing, 
the International AGN Watch is completing preliminary analysis of
49 days of nearly continuous observations of NGC 7469 with \IUE, one of the
high-priority ``lasting value'' projects undertaken during its 
nineteenth and final year of operations, which produced somewhat more than 200
independent spectra (Wanders et al.\ 1997). Intensive observations were made
simultaneously with ground-based telescopes (Collier et al.\ 1997)
and with the {\it Rossi X-Ray Timing Explorer (RXTE)}\/
(Nandra et al.\ 1997).
Light curves for the UV continuum bands and emission
lines are shown in Fig.\ 4, along with the cross-correlation
functions. 
\begin{figure}
\epsfxsize\hsize
\epsfbox{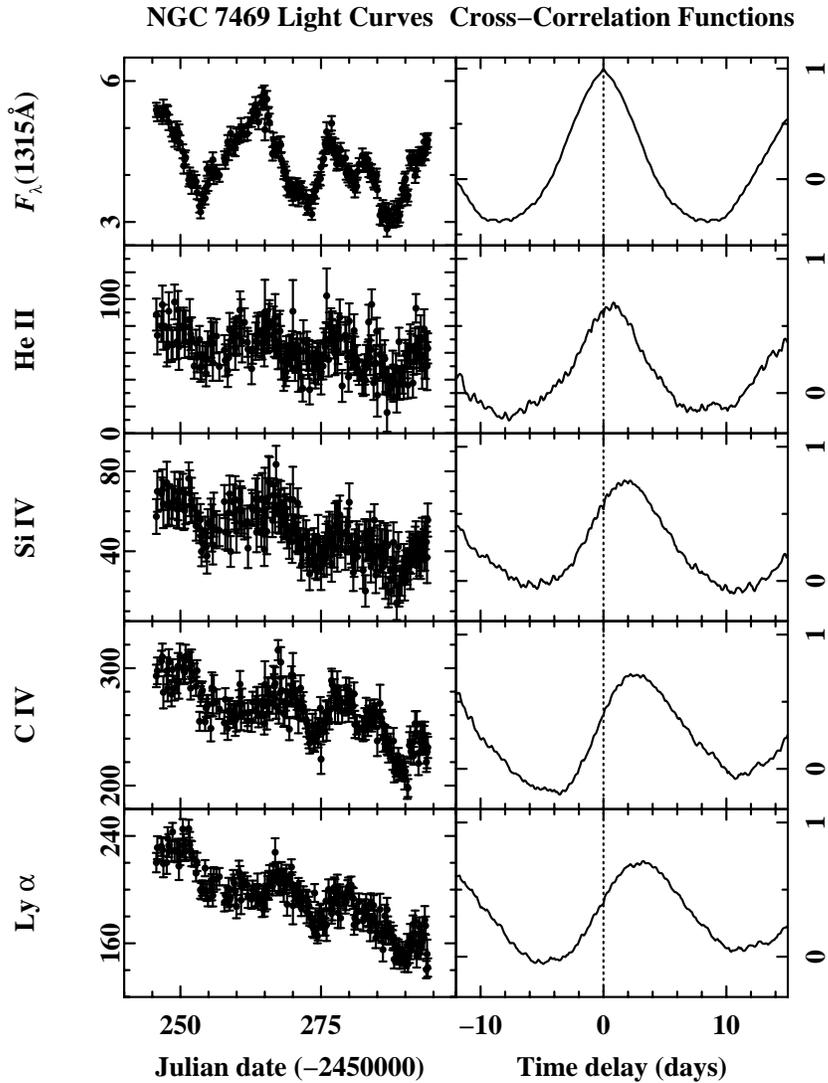}
\caption{The left-hand columns show the light curves of NGC~7469
obtained with \IUE\ during an intensive AGN Watch monitoring
campaign during the summer of 1996. The right-hand column shows
the result of cross-correlating the light curve immediately to
the left with the 1315\,\AA\ light curve at the top of the
left column; the panel at the top of the right column thus shows
the 1315\,\AA\ continuum autocorrelation function.
Data from Wanders et al.\ (1997).}
\end{figure}

\section{Evaluation of Reverberation-Mapping Results}

\subsection{The gas distribution in the BLR}
In this section, we attempt to answer the fundamental questions of 
reverberation mapping by evaluating the results of the more successful
International AGN Watch campaigns. 
We address, in more detail, several of the uncertainties,
 and proceed to evaluate the significance
of the available 
lag determinations. We then discuss the gas distribution in the BLR, 
as obtained by direct and
indirect (inversion) methods.

\subsubsection{Limitations}
Obtaining well-sampled line and continuum light curves that
are relatively noise free is only the first step towards
 the goal of a complete mapping of the  gas distribution in the BLR. The 
interpretation of the data is not straightforward and 
there are various   ambiguities  and  uncertainties. Three of the more
 critical problems are explained below.
\begin{enumerate}
\item
A very severe limitation is the non-linear response of many emission lines.
All lines  are produced in a restricted zone  inside the cloud  and the line
emissivity is a sensitive function of the incident ionizing flux, 
i.e.,  of the ionization parameter
\begin{equation}
 U = \frac{Q}{4 \pi R^2 n_{\rm H} c},
\end{equation}
where $Q$ is the rate at which the source produces
ionizing photons (i.e., photons per second). Consider as an
example \CIV, whose emission reflects both
the size of the C$^{+3}$ zone and the gas temperature over this region.
 This line flux increases
linearly with the continuum flux only 
if most carbon is C$^{+2}$ or C$^{+3}$. Further 
increase in the continuum level makes C$^{+4}$
 the dominant species and results in a large
decrease in the emissivity of the line. The transfer function of \CIV\ 
 reflects
the non-linear nature of the ionization process and the corresponding
change in temperature. Most metal lines behave in a similar way, but
over a different ionization-parameter range. 
The only exception is \La, whose response
is close to linear under most conditions,
 since the line intensity directly
reflects the number of ionizing photons absorbed by the gas.
The  \La\  line reacts non-linearly to the
incident ionizing flux in two extreme cases; 
in low column-density clouds, 
the increased continuum level can make the gas fully  transparent, 
thus reducing the line production considerably. In
high column-density clouds, continuum variation can cause a significant 
change in the line optical depth $\tau($\La) 
which will reduce the line emissivity due to line
trapping and collisional suppression. Thus, non-linear effects are of great
significance and affect the shape of the transfer function.
Figure 5 illustrates the 
result of a particular set of photoionization calculations applied to the 
case of NGC~5548, the best observed Seyfert 1 galaxy. The model  assumes
a specific gas distribution given by
$\nh\propto  R^{-1}$, which results in a
radial dependence of the ionization parameter,
$U\propto  R^{-1}$.
The column
density  is obtained by assuming spherical clouds that retain their identity
as they move in or out at their virial speed (for more details see 
Netzer 1990). 
The continuum luminosity is scaled to the
luminosity of NGC~5548 during the 1989 campaign. With the measured 
\Lya\ lag of about 9 days,
and the required, radius-averaged ionization parameter 
obtained from  observed line ratios, this leaves little 
freedom for the choice of density at \Rin.
\begin{figure}
\epsfxsize\hsize
\epsfbox{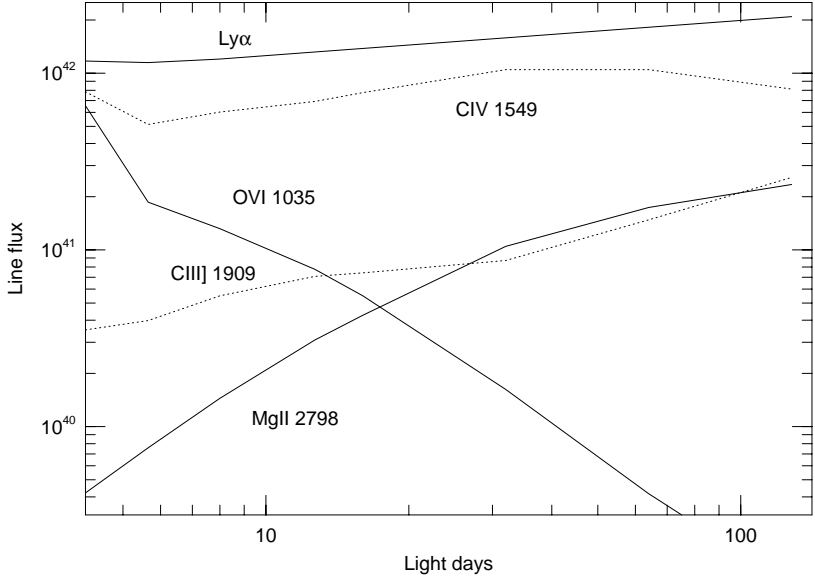}
\caption{Line luminosity, per unit solid angle, as a function of 
distance from the center of NGC~5548 for an assumed gas distribution of
$\nh \propto R^{-1}$ and  $\Ncol\propto R^{-2/3}$.}
\end{figure}
The diagram illustrates the constancy of some lines (\La\ and \CIV) 
and the dependence
of other lines on the radius and ionization parameter 
 (\OVI\ and \MgII) and the density (\CIIIb).
\item
A second major source of uncertainty is the unknown shape and variability 
amplitude of the  ionizing continuum.
Multi-wavelength spectroscopy suggests that the shape of the 
optical--UV continuum changes
with luminosity. It is thus reasonable to assume that the Lyman continuum,
which drives the line luminosity, is changing its shape too. Present 
techniques  assume
that the variability of the ionizing flux is similar to the  
observed variability at
much longer wavelengths, thus introducing an uncertainty 
into the transfer function.
In principle, the shape of the ionizing continuum 
can be deduced from various
line ratios; in practice, this is a difficult task. 
\item
Line emission from the BLR is not necessarily isotropic. The most notable 
example is \La, whose emission pattern is likely to be beamed towards the
central source because of the large column of neutral gas at the
back of BLR clouds. A known beaming pattern can, in principle, be taken
into account when calculating the transfer function 
(see Ferland \etal\ 1992),
but present photoionization models cannot reliably calculate this pattern.
\end{enumerate}

\subsubsection{Lag determination}

Even the simplest of all reverberation-mapping aims, 
the lag determination, is not 
a simple task since, in many cases, the line-to-continuum 
lag is only a small fraction of the typical variability time scale. 
Medium-luminosity Seyfert 1s are characterized by 
variability time scales of 30--100
days and lags of 2--6 days for the high-ionization lines.
Since the lag is determined from the peak (or the center
of gravity) of the CCF, which is the convolution of the ACF with the transfer
function (eq.\ \ref{eq:CCF}), 
the required accuracy is a small fraction of the CCF width.
As noted earlier,
there are several methods of determining the CCF, 
but none with a rigorous way of estimating the associated uncertainty.
Thus, a 
slight misjudgment of the center of the asymmetric CCF 
gives a large uncertainty on the lag. 

A simple, yet powerful method to determine the uncertainty associated
with the lag  is via Monte-Carlo simulations. 
This is done by guessing a plausible BLR model, computing
its response to a given input signal, and then sampling
the input and response light curves in a fashion similar to 
what was achieved in the actual experiment, including the
effects of random and, in some cases, systematic errors.
The assumed known
parameters are the overall gas distribution (or, more precisely, the
emissivity distribution of a particular line) and the driving continuum light
curve. The observed continuum light curve is interpolated, in a predetermined
way, to obtain the ``known'' continuum behavior. 
This smooth light curve is then fed
through a pre-determined geometry, e.g., a thick shell with known \Rin\
and \Rout\, and emissivity as a function of radius, to obtain a 
predicted ``line
light curve''. This line light curve is  sampled in a pattern similar to the
actual observations.
A noise model is added to the ``observations'', and the
cross-correlation is performed, just as if these were real data,
to obtain a CCF and its peak or centroid location.
Many repetitions of this process produce
a cross-correlation peak distribution (CCPD; Maoz \& Netzer 1989)
which gives, for the chosen
geometry and observing pattern, the probability of finding the peak of
the CCF inside a certain interval, say 3--5 days. This is probably the best 
way of assessing the reality of a particular lag determination 
and its associated uncertainty, for a given geometry. 
Figure 6 shows a specific
example for a geometrically thick BLR in a luminous 
($L=2\times10^{45}$\,\ergsec) quasar.
\begin{figure}
\epsfxsize\hsize
\epsfbox{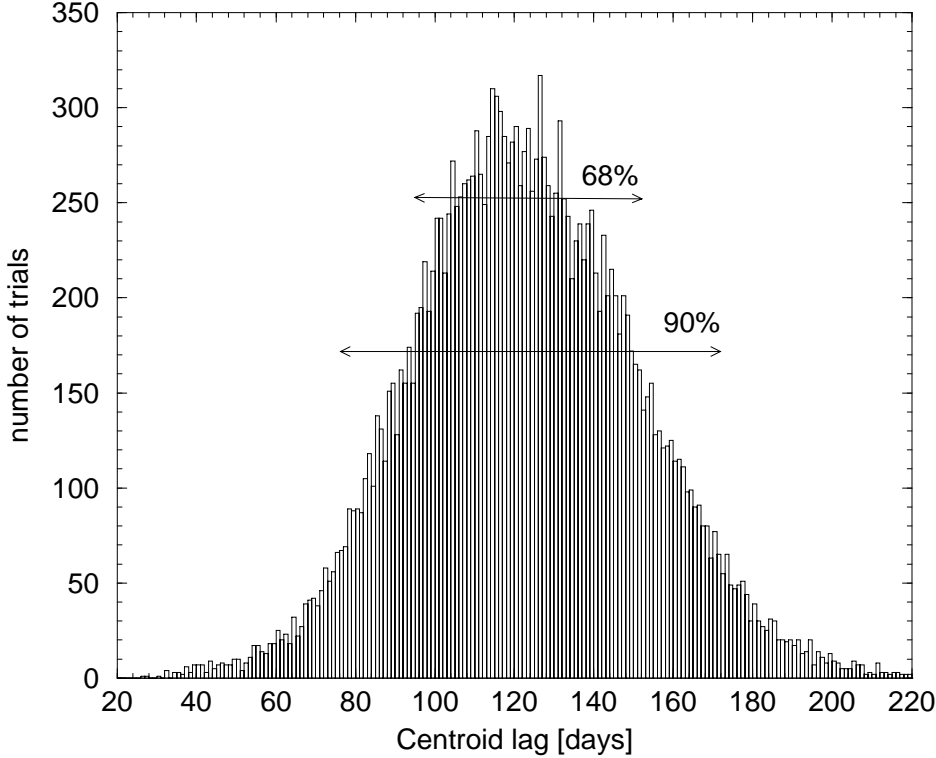}
\caption{The CCPD histogram for the quasar
PG 0804$+$762 (for the light curve see Kaspi \etal\ 1996b) assuming
a thick spherical shell, extending from 20 to 200 light days, with 
line emission proportional to $dR$. The 68\% and 90\% intervals
deduced from the simulations are shown.}
\end{figure}
The principal weakness of this method is that the results are only
valid to the extent that the model is a reasonable representation of
reality. Monte-Carlo methods require a complete
set of pre-determined
assumptions and are very useful in those cases 
where the number of assumptions
is small and the line response to continuum variability can be 
reliably simulated.

\subsubsection{Gas distribution using inversion methods}

Attempts to recover the one-dimensional transfer
function (eq.\ \ref{eq:1dTF}) have been made for several sources, including
NGC 4151 (Maoz et al.\ 1991),
NGC 5548 (Horne, Welsh, \& Peterson 1991; Krolik et al.\ 1991;
Peterson et al.\ 1994; Wanders \& Peterson 1996),
Mrk 590 (Peterson et al.\ 1993),
NGC 3516 (Wanders \& Horne 1994), and
NGC 3227 (Winge et al.\ 1995). In general, the data are insufficient
to show any clear structure in $\Psi(\tau)$, although in many cases
there is some evidence for very weak line response at $\tau \approx 0$,
i.e., due to material along our line of sight to the continuum source. 
It has been suggested (Ferland \etal\ 1992; O'Brien, Goad, \& 
Gondhalekar 1994) 
that this might be due to emission-line anisotropy, as
explained earlier.
 There are still doubts,
however, about whether or not the lack of line response at zero lag
is in fact real (see Maoz 1997). 

MEM methods are somewhat limited and, as
shown by Horne (1994),  can produce several different-looking transfer
functions that fit the data equally well. The uncertainty in the width of
such transfer function is large, due to the broad intrinsic ACF.
 In some cases the differences between acceptable solutions
are large enough to prevent us from reaching even the simplest conclusions
regarding the response at zero lag.
The requirements on the amount and
quality of the light-curve data are sufficiently 
demanding that we doubt that anyone
would claim that today a unique and unambiguous solution has been found 
using inversion methods. 

In only two cases have attempts been made at
full recovery of $\Psi(v,\tau)$, in both cases for 
the \civ\,$\lambda1549$ emission line, (1) using
\HST\/ spectra of NGC 5548 (Wanders et al.\ 1995; Done \& Krolik 1996), and
(2) using \IUE\/ spectra of NGC 4151 (Ulrich \& Horne 1996;
see also Horne's paper in this volume). In both cases, predominantly
radial gas motions seem to be ruled out by the similarity of the response
of the redshifted and blueshifted sides of the emission line. However,
in both cases, there is quite a bit of ambiguity about the results,
and the transfer-function structure is actually poorly determined.
In the case of NGC 5548, the structure seems broadly consistent with the
model shown in Fig.\ 3, as mentioned above (see Wanders et al.\ 1995).
However, the details are not matched well by the model. Indeed there
is a possible indication of an infalling component (Done \& Krolik 1996)
based on more rapid response of the red side of the line profile compared to 
the blue side, but which Wanders et al.\ (1995) dismiss as
probably stochastic (i.e., the apparent ``direction of flow''
varies with time). Ulrich \& Horne (1996) also suggest that there is
a weak infall component present in the NGC 4151 data, but again,
the signature is not unambiguous.

\subsubsection{Gas distribution using direct (forward) methods}
 
Inversion methods are designed to produce emissivity maps for
specific lines while the physical picture we are after 
involves the run of density, column density, 
and cloud properties across the BLR.
Unfortunately, there are large uncertainties in converting from emissivity maps
to the desired physical maps, all associated with our limited understanding of
the line production mechanism. The result is that, 
so far, there is not a single published model that can 
explain the light curves of all observed emission lines even in the
best-studied AGN. 
It is therefore important to search for a 
simpler way of obtaining reliable information on 
some of the important parameters. This is provided by direct methods.

Direct methods  make assumptions about the gas distribution 
and other properties and  calculate, under these assumptions, the
line intensity at every point and for every continuum flux level.   
The integrated line fluxes, as a function
of time, are then compared with the observations.  
The procedure is easily followed in the particular case where
all important
properties, such as the gas density and column density, are given by smooth
simple functions of  the distance from the center 
(Netzer 1990). In such cases, the dependence of line
emissivity on distance can be  calculated by photoionization models and
the procedure repeated 
a number of times with various assumed  geometries.
Final decision on the best chosen dependence is based on the model which
best reproduces all of the emission lines.
 
Figure 7 illustrates two results of direct calculations
applied to the   \La\ light curve in the specific case of NGC~5548
discussed earlier (Fig.\ 5). In this case, all parameters are fixed by the
assumed run of \nh\ with distance and the only missing assumptions are
the values of \Rin\ and \Rout. The diagram compares the observed
light curve with the model predictions for \Rin=4 light days and two
choices of \Rout, 30 light days and 100 light days. 
For both of these we note
the predicted lag and the quality of the fit ($\chi^2$). Note that the
assumed observational uncertainties are artificially increased,
 beyond those observed, 
for earlier times, to allow for the large uncertainty in the beginning of
the campaign (see \S2.1). Thus, at day 1, they are three times larger than
the actual observed uncertainties (8\% in this case). 
\begin{figure}
\epsfxsize\hsize
\epsfbox{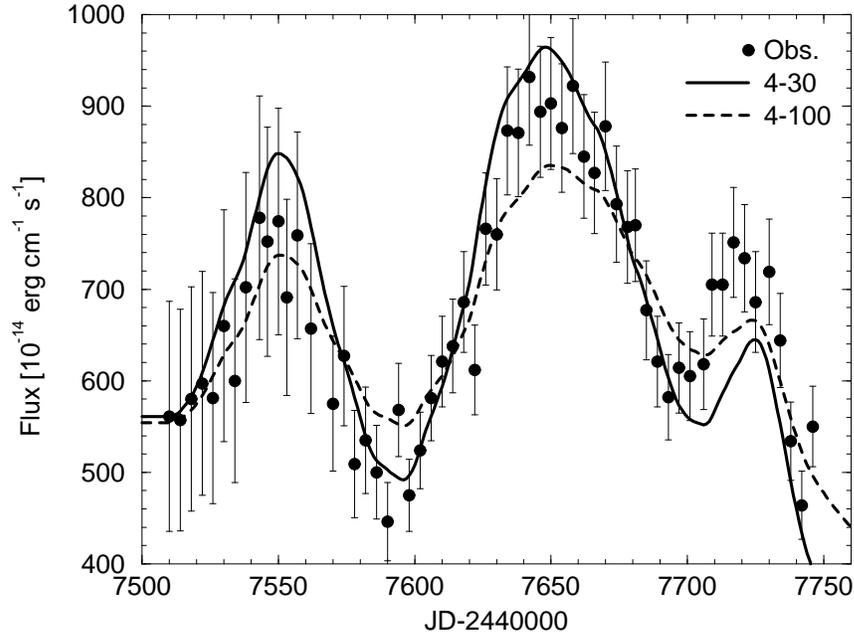}
\caption{The observed \La\ light curve, from the 1989 NGC~5548 campaign,
is compared with two predicted geometries involving $\nh\propto R^{-1}$.
In both cases $\Rin = 4$ light days and two different values of
\Rout, 30 and 100 light days. The value
of $\chi^2$ for 55 degrees of freedom is 75 for the small geometry
and 54 for the large geometry. This $\chi^2$ has been
calculated with the shown 
error bars that have been artificially increased, for days 1--80, to allow
for the unknown continuum behavior before the beginning of the campaign.}
\end{figure}
The diagram illustrates that the assumed density dependence is close, but not
in perfect agreement with the observations. Better modeling
requires somewhat different radial density dependence. The reality of
such models should be assessed by checking the prediction for
other emission-line light curves.

Diagrams like Fig.\ 7 suffer from the well-known non-linear response which
is a problem in all lines, especially metal lines. There is, however,  one
measured quantity that is almost independent of this,  the {\it total line
emission} (Netzer 1990). Simple energy conservations arguments suggest
that the total intensity of all emission lines equals the ionizing energy
absorbed by the cloud. Summing up the line intensities  gives the
almost-perfect and easiest-to-model light curve. This is  ``almost perfect''
since the above sum involves some lines that are outside the observable
range and some emission (mostly bound--free) that is difficult to measure.
However, it is superior to all other methods and enables us, with some simple
assumptions, to obtain a reliable estimate on the gas distribution. Robust
direct methods are those that screen, out of all possible geometries, only
those that are consistent with the ``total line emission'' light curve.
The only major unknown in this procedure is the one associated with
anisotropic line emission.    

\subsection{The ``New'' BLR}

Ten years of monitoring campaigns 
have resulted in a clearer view of
AGN physics and in better understanding of the BLR. Inversion
methods have provided some insight into the gas distribution and dynamics,
but fall short of resolving the more fundamental
issues. Direct methods and lag determinations have resulted
in some very meaningful new results and
we feel it is time to  compare the state of the field today to what it was
 ten years ago:
\begin{enumerate}
\item
The size of the ``old'' BLR was determined indirectly from the 
best-estimated ionization parameter 
by making assumptions about the incident ionizing flux
and the gas density. 
The 1997 BLR size is directly measured and is  much smaller, by almost an 
order of magnitude. 
\item
The 1987 BLR was assumed to be ``thin'', with $\Rin\approx\Rout$. The
1997 BLR is very thick, with $\Rin/\Rout \leq 0.1$. This is clearly
demonstrated by the variable measured lags, in a single source, as a function
of the continuum event duration (Netzer \& Maoz 1990). It is verified by the
shape of the derived transfer functions and by direct modeling.
\item
When many lines have been observed in a single source, there is clear
evidence for radial ionization stratification of the BLR, as the 
highest-ionization lines respond more rapidly than the lower-ionization lines.
In NGC 5548, for example, \nv\,$\lambda1240$ and
\heii\,$\lambda1640$ show lags of 1--2 days,
\Lya\ and \civ\ have lags of 7--12 days, and
\Hbeta\ responds even more slowly. 
\item
There is {\it some}\/ evidence
that the lag changes with time, at least in 
NGC 5548 (Peterson et al.\ 1994); \Hbeta\ lags ranging from 
7 days (Netzer et al.\ 1990)
to 22 days (Peterson et al.\ 1994) have been reported for this source,
and indeed, the quantity we call the lag can vary on surprisingly
short time scales.
\item
 The gross similarity of
AGN spectra over a wide range in luminosity led, more
than ten years ago, to a na\"{\i}ve
prediction that the ionization parameter is almost independent of
luminosity and thus $\RBLR \propto L^{1/2}$.
Today we are seeing the first proof of this theoretical prediction
through direct measurements of \RBLR\ in high-luminosity sources. 
\end{enumerate}

\subsection{Black-hole mass and the nature of the powerhouse}
Determining the black-hole mass was the most important motivation
for all of the various AGN Watch monitoring campaigns. 
Such a determination must assume some kind
of virial motion that is  experimentally not yet proven.
A key issue  is the best combination
of the BLR size and the gas velocity. The CCF peak, or the transfer function,
provides an emissivity-weighted radius and, without a full dynamical
model, it is not simple to choose the corresponding velocity. So far, the 
full-width at half maximum (FWHM)
has been used in combination with this peak, but there are reasons to prefer 
the minimum inferred size \Rin\  combined with the maximum observed velocity
as obtained from the emission-line 
FWZI. Sharply peaked transfer functions, like those obtained 
for high-excitation lines, give the best estimate for 
\Rin.
The uncertainty associated with the mass determination, due to the
ambiguous size--velocity combination, is around a factor of two. Needless to
say, the relative mass of different objects must be based on the same size
determination, i.e., the transfer function or the CCF peak as determined
for  the same emission lines.

It is only recently that serious attempts to probe the  
nature of the powerhouse itself have been made. The common
wisdom is that the AGN continuum is produced 
in close proximity to the central black hole,
in a thin or slim disk. The X-ray continuum is produced in the innermost
part of this structure while the optical-UV radiation is produced farther
out. The two main ideas about the UV radiation involve the release
of gravitational energy in the disk or the reprocessing of the central
X-ray radiation (which, in itself, must be driven by accretion).
A main problem is the energy budget since it is not obvious that the
available X-ray energy can provide the energy observed in the
UV flares.

Interestingly, the above two possibilities are associated
with very different time scales. The viscous time determines the disk 
reaction to a change in accretion rate while the light-travel time gives
the typical scale associated with X-ray reprocessing. Measuring the lag (if
any) between X-ray and UV events could solve this issue.
Using the standard thin disk assumptions we note that for such systems
\begin{equation}
T_{\rm eff}^4 = \frac{3GM \dot{M} }{8 \pi \sigma R^3} 
\left[ 1- \left( \frac{R}{R_{\rm ms}} \right)^{-1/2} \right],
\end{equation}
where $R_{\rm ms}$ is the radius of marginal stability
($6R_{\rm g}$ for Schwarzschild black hole and 
$1.22R_{\rm g}$ for a maximally
rotating Kerr hole) and we have neglected relativistic corrections.
Assuming for simplicity that the frequency of the peak blackbody emissivity,
$h \nu_{\rm max} = 2.8kT$, 
represents the emission at every frequency, and adopting the simple 
$R\propto T^{-4/3}$ dependence, we get 
$R\propto \nu_{\rm max}^{-4/3}$, or, using $r = R / R_{\rm g}$,
\begin{equation}
r \approx 1.8 \times 10^{22}\, \nu_{\rm max}^{-4/3}\,
M_8^{-2/3} \dot{M}_{\odot}^{1/3}, 
\end{equation}
where $M_8 = M/10^8 M_{\odot}$ and $\dot{M}_{\odot}$ is
the accretion rate in solar masses per year. 
This couples the size and the location corresponding to a given
maximum frequency of a given luminosity AGN. For example, the best values
obtained for  NGC~5548 
($M_8 \approx 0.5,  \dot{M}_{\odot} \approx 0.07$) combined with 
 a Schwarzschild black hole, gives a separation  between the
optical (5000\,\AA) and the UV (1400\,\AA) peak emission 
locations that corresponds to about 0.6 light days. 
This spacing suggests a  short light-travel time and
a very long 
viscous time for the correlated UV and optical continuum variability in
this source. The light-travel time is consistent with
the measured upper limit (less than two days; see Korista et al. 1995).

Prior to the recent NGC~7469 campaign, there was no real measurement of 
optical--UV or UV--X-ray continuum lags. This is likely to change
soon and the results will definitely shed new light on this issue and
may enable us, for the very first time, to distinguish between the simple
disk model and the reprocessing idea.
 
\subsection{Luminosity--size--mass relationship for AGN}
One of the  exciting results reported in this meeting is the
first meaningful size--luminosity relationship for AGN 
(Kaspi \etal\ 1996b, and Kaspi's contribution to these proceedings). 
For the first time, we have 
a large enough AGN sample covering almost two orders of 
magnitude in luminosity and
showing a clear trend of  \Hb\ lag as a function of luminosity. The 
best-fitted lag--luminosity slope is very close to the predicted 
$R\propto L^{1/2}$ relationship, and the resulting size is approximately
\begin{equation}
R_{\rm BLR} = 0.014 L_{44}^{1/2}\;{\rm pc}
\end{equation}
where
$L_{44}$ is the 0.1--1\mic\ luminosity in units of 10$^{44}$\,\ergsec\
(we have assumed $\Ho = 75\,\Hubble$ and $\qo = 0.5$).

The search for an $L$--$M$ 
relationship involves yet another correlation of $L$ with the emission-line
FWHM. 
Unfortunately, the sample used for determining the $L$--size correlation 
is too small and the spread in line width  too weak to
use it for this purpose. 
Reports in the literature involving much larger samples
suggest anything from 
FWHM independent of $L$  to 
${\rm FWHM} \propto  L^{1/4}$. Obviously, strong 
selection effects that are not well understood are involved. 
It is also becoming
apparent that the broad-line widths depend on  
X-ray properties (Boller \etal\ 1996) 
and that different lines can show different widths and dependences. 
Given all this, and the experimentally verified 
$R_{\rm BLR} \propto L^{1/2}$, 
it is probably safe to say that the  black-hole
mass determination is consistent 
with anything from  $M\propto L$
to $M\propto L^{1/2}$, with a factor 2--4  uncertainty
on the normalization of this relationship.

\section{Future Reverberation-Mapping Experiments}
The last 10 years were so full of surprises that we shall not risk predicting
the 10 coming ones. Instead, we  comment on what can be done, and should be
tried, with present-day capabilities.

\subsection{Improved sampling and multi-wavelength campaigns}
Improved sampling has already resulted in some spectacular results. The recent
dense sampling of NGC~4151 (Crenshaw \etal\ 1996, Kaspi \etal\ 1996a) has shown
continuum variability on extremely short time scales,
and the soon-to-be-published
results of the  NGC~7469 monitoring (Fig.\ 4)
will no doubt raise some new questions about the origin of the
UV and X-ray continuum.

Unfortunately, there is not much hope for repeating the long-duration (many
months), dense-sampling campaigns like those that involved the late \IUE.
\HST\ observations can provide superior data
that are extremely important for continuum measurements but of 
insufficient duration for dense sampling of 
emission lines. Some X-ray satellites, like 
{\it RXTE}, are suitable for 
dense, high-frequency sampling,
but it seems unlikely that {\it AXAF}\/ or {\it XMM}\/ will spend such long
times on individual targets. 
  
Multi-wavelength campaigns are
likely to continue to be the most important way of probing the 
physics of the BLR, with important implications for the 
continuum-emitting source. The combination of X-ray information with data at
lower photon energies is crucial. While ground-based telescopes are 
likely to contribute a lot in this area, 
the weak link is the UV, which now depends almost solely on \HST. Long-term
\HST\/ programs are definitely necessary to proceed in this direction.

It is important to note that {\it SRG}\/ (to be launched
at the end of 1998) will carry on board two X-ray telescopes 
{\it (SODART} and {\it JET-X)}
and a UV telescope {\it (TAUVEX)} 
that are capable of producing the required long-term 
multi-wavelength continuum monitoring. No coordination is required in 
this case since all telescopes on board this mission are looking at the same
field. It remains to be seen whether the observing program of the satellite
will allow the very long observations required for such projects. 

\subsection{Large-area, multi-object campaigns}
Most AGN campaigns, so far, have concentrated on  detailed studies of 
individual objects. The result is a small sample covering a limited range in
luminosity and redshift. Very high-luminosity,
high-$z$ quasars are completely missing.

Future campaigns will be able to use multi-aperture fiber-fed spectrographs
to improve on this situation. Such a technique 
will allow repetitive observations
of several dozens of AGN, in the same field, for a long period of time. 
Infrequent visits, e.g., every couple of weeks, are definitely adequate to
secure a large sample of high-$z$ quasars and to measure 
their BLR sizes. Given the predicted values of \RBLR\  and the 
high redshifts,
such a campaign would probably require 5 to 10 years.
  
\subsection{Future role of the Wise Observatory and other small telescopes}
While future spectroscopy of large, high-$z$ AGN samples is limited to large 
telescopes, small telescopes are still very important. They can provide
 frequent, broad-band imaging of the same fields thus adding  many more
points to the continuum light curves. As shown in 
many previous campaigns, adequate sampling of
the continuum light curve is a critical issue. Emission lines react slowly
to continuum variations and their sampling, 
given a well-constrained continuum,
is less crucial. 
Small telescopes with large fields of view, like the 1-m telescope at the Wise
Observatory, can easily provide  high-frequency continuum 
observations of luminous AGN to
supplement the spectroscopic observations  by the large telescopes.
\vspace*{0.5cm}

BMP gratefully acknowledges support for reverberation-mapping studies
at Ohio State University by the US National Science Foundation through
grant AST-9420080 and by NASA through Long-Term Space Astrophysics Grant
NAG5-3233 and Astrophysics Data Program Grant NAG5-3497.
Monitoring of AGN at the Florence and George Wise Observatory is supported
by the Israel Science Foundation 
and by the Raymond and Beverly Sackler Institute of Astronomy.
\bigskip

\noindent{\bf References}
\medskip

\leftskip=1em
\parindent=-1em

\footnotesize
Alloin, D., Clavel, J., Peterson, B.\,M., Reichert, G.\,A., \& 
Stirpe, G.\,M. 1994, in {\it Frontiers of Space 
and Ground-Based Astronomy}, ed.\ 
W.\,Wamsteker, M.\,S. Longair, \& Y.\ Kondo 
(Dordrecht: Kluwer), p.\ 423

Alloin, D., et al. 1995, A\&A, 293, 293

Blandford, R.\,D., \& McKee, C.\,F. 1982, ApJ, 255, 419

 Boller, Th., Brandt, W.N, \& Fink, H. 1996, A\&A, 305, 53 

 Carone, T.\,E., et al. 1996, ApJ, 471, 737

 Clavel, J., et al. 1991, ApJ, 366, 64

 Collier, S., et al. 1997, in preparation

 Crenshaw, D.\,M., et al. 1996, ApJ, 470, 322

 Dietrich, M.,  et al. 1993, ApJ, 408, 416

 Dietrich, M.,  et al. 1997, in preparation

 Done, C., \& Krolik, J.\,H. 1996, ApJ, 463, 144

 Edelson, R.,  et al. 1996, ApJ, 470, 364

 Ferland, G.\,J., Peterson, B.\,M., Horne, K., Welsh, W.\,F., 
\& Nahar, S.\,N. 1992, ApJ, 387, 95

 Goad, M., \& Wanders, I. 1996, ApJ, 469, 113

 Horne, K. 1994, in {\it Reverberation Mapping of the Broad-Line
Region in Active Galactic Nuclei}, ed.\ P.\,M.\ Gondhalekar, K.\ Horne, \&
B.\,M.\ Peterson, ASP Conference Series Vol.\ 69 (San Francisco: ASP), p.\ 23

 Horne, K., Welsh, W.\,F., \& Peterson, B.M. 1991, ApJ, 367, L5

 Kaspi, S., et al. 1996a, ApJ, 470, 336

 Kaspi, S., Smith, P.\,S., Maoz, D., Netzer, H., \&
Jannuzi, B.\,T. 1996b, ApJ, 471, L75

 Kassebaum, T.\,M., Peterson, B.\,M., Wanders, I., Pogge, R.\,W.,
Bertram, R., \& Wagner, R.\,M. 1997, ApJ, 475, 106

 Korista, K.\,T.,  et al. 1995, ApJS, 97, 285

 Krolik, J.\,H., \& Done, C. 1995, ApJ, 440, 166

 Krolik, J.\,H., Horne, K., Kallman, T.\,R., Malkan, M.\,A.,
Edelson, R.\,A., \& Kriss, G.\,A. 1991, ApJ, 371, 541

 Leighly, K.\,M., O'Brien, P.\,T., Edelson, R.,  George, I.\,M.,
Malkan, M.\,A., Matsuoka, M., Mushotzky, R.\,F., \& Peterson, B.\,M. 1997, 
ApJ, in press 

 Maoz, D. 1997, in {\it Emission Lines in Active Galaxies: New
Methods and Techniques}, ed.\ B.\,M.\ Peterson, F.-Z.\ Cheng, 
\& A.\,S.\ Wilson, ASP Conference Series Vol.\ 113 
(San Francisco: ASP), p.\ 138
 
Maoz, D., \& Netzer, H. 1989, MNRAS, 236, 21 

 Maoz, D., Netzer, H., Leibowitz, E., Brosch, N., Laor, A.,
Mendelson, H., Beck, S., Almoznino, E., \& Mazeh, T. 1991, ApJ, 367, 493

 Maoz, D., Netzer, H., Mazeh, T., Beck, S., Almoznino, E.,
Leibowitz, E., Brosch, N., Mendelson, H., \& Laor, A. 1990, ApJ, 351, 75

 Marshall, H.\,L., Carone, T.\,E., Peterson, B.\,M., 
Clavel, J., Crenshaw, D.\,M., Korista, K.\,T., Kriss, G.\,A.,  Krolik, J.\,H.,
Malkan, M.\,A.,  Morris, S.\,L., O'Brien, P.\,T., \& Reichert, G.\,A. 1997,
ApJ, 479, 222

 Nandra, K., et al. 1997, in preparation

 Netzer, H. 1990, in {\it Active Galactic Nuclei}, 
R.D.\ Blandford, H.\ Netzer, and L.\ Woltjer (Berlin: Springer-Verlag),
p.\ 57

  Netzer, H., \& Maoz, D. 1990, ApJ, 365, L5

 Netzer, H., Maoz, D., Laor, A., Mendelson, H.,
Brosch, N., Leibowitz, E., Almoznino, E., Beck, S.,  
\& Mazeh, T. 1990, ApJ, 353, 108

 O'Brien, P.\,T., Goad, M.\,R., \& Gondhalekar, P.\,M. 1994,
MNRAS, 268, 485

 O'Brien, P.\,T., et al. 1997, in preparation

 Penston, M.\,V. 1991, in {\em Variability of Active Galactic Nuclei},
ed.\ H.R.\ Miller \& P.J.\ Wiita
(Cambridge, Cambridge Univ.\ Press), p.\ 343

 P\'{e}rez, E., Robinson, A., \& de la Fuente, L. 
1992, MNRAS,  256, 103 

 Peterson, B.\,M. 1988, PASP, 100, 18

 Peterson, B.\,M. 1993, PASP, 105, 247

 Peterson, B.\,M., et al.\ 1991, ApJ, 368, 119

 Peterson, B.\,M., et al.\ 1992, ApJ, 392, 470

 Peterson, B.\,M., et al.\ 1993, ApJ, 402, 469

 Peterson, B.\,M., et al.\ 1994, ApJ, 425, 622

 Peterson, B.\,M., et al.\ 1997, in preparation

 Pijpers, F.\,P., \& Wanders, I. 1994, MNRAS, 271, 183

 Reichert, G.\,A., et al.\ 1994, ApJ, 425, 582

 Robinson, A. 1994, in {\it Reverberation Mapping of the Broad-Line
Region in Active Galactic Nuclei}, ed.\ P.\,M.\ Gondhalekar, K.\ Horne, \&
B.\,M.\ Peterson, ASP Conference Series Vol.\ 69 (San Francisco: ASP), p.\ 147

 Rodr\'{\i}guez-Pascual, P.\,M., et al.\ 1997, ApJS, in press

 Romanishin, W., et al. 1995, ApJ, 455, 516

 Santos-Lle\'{o}, M., et al.\ 1997, ApJS, in press

 Stirpe, G.\,M., et al. 1994, ApJ, 425, 609

 Ulrich, M.-H., \& Horne, K. 1996, MNRAS, 283, 748

 Wanders, I., 
et al.\ 1995,  ApJ, 453, L87

 Wanders, I., \& Horne, K. 1994, A\&A, 289, 76

 Wanders, I., \& Peterson, B.\,M. 1996, ApJ, 466, 174

 Wanders, I., et al. 1997, ApJS, in press

 Warwick, R., et al. 1996, ApJ, 470, 349

 Welsh, W.\,F.,\& Horne, K. 1991, ApJ, 379, 586

 White, R.\,J., \& Peterson, B.\,M. 1994, PASP, 106, 879

 Winge, C., Peterson, B.\,M., Horne, K., Pogge, R.\,W., Pastoriza,
M.\,G., Storchi-Bergmann, T. 1995, ApJ, 445, 680 
 
Winge, C., Peterson, B.\,M., Pastoriza, M.\,G., 
Storchi-Bergmann, T. 1996, ApJ, 469, 648

\end{document}